# MAORY: A Multi-conjugate Adaptive Optics RelaY for ELT

Paolo Ciliegi[1]
Guido Agapito[1]
Matteo Aliverti[1]
Francesca Annibali[1]
Carmelo Arcidiacono[1]
Andrea Balestra[1]
Andrea Baruffolo[1]
Maria Bergomi[1]
Andrea Bianco[1]
Marco Bonaglia[1]
Lorenzo Busoni[1]
Michele Cantiello[1]
Enrico Cascone[1]
Gaël Chauvin[3]
Simonetta Chinellato[1]
Vincenzo Cianniello[1]
Jean-Jacques Correia[3]
Giuseppe Cosentino[1]
Massimo Dall'Ora[1]
Vincenzo De Caprio[1]
Nicholas Devaney[2]
Ivan Di Antonio[1]
Amico Di Cianno[1]
Ugo Di Giammatteo[1]
Valentina D'Orazi[1]
Gianluca Di Rico[1]
Mauro Dolci[1]
Sylvain Doutè[3]
Cristian Eredia[1]
Jacopo Farinato[1]
Simone Esposito[1]
Daniela Fantinel[1]
Philippe Feautrier[3]
Italo Foppiani[1]
Enrico Giro[1]
Laurance Gluck[3]
Aaron Golden[2]
Alexander Goncharov[2]
Paolo Grani[1]
Marco Gullieuszik[1]
Pierre Haguenauer[4]
François Hénault[3]
Zoltan Hubert[3]
Miska Le Louran[4]
Demetrio Magrin[1]
Elisabetta Maiorano[1]
Filippo Mannucci[1]
Deborah Malone[2]
Luca Marafatto[1]
Estelle Moraux[3]
Matteo Munari[1]
Sylvan Oberti[4]
Giorgio Pariani[1]
Lorenzo Pettazzi[4]
Cédric Plantet[1]
Linda Podio[1]
Elisa Portaluri[1]
Alfio Puglisi[1]
Roberto Ragazzoni[1]
Andrew Rakich[1]
Patrick Rabou[3]
Edoardo Redaelli[1]
Matt Redman[2]
Marco Riva[1]
Sylvain Rochat[3]
Gabriele Rodeghiero[1]
Bernardo Salasnich[1]
Paolo Saracco[1]
Rosanna Sordo[1]
Marilena Spavone[1]
Marie-Hélène Sztefek[3]
Angelo Valentini[1]
Eros Vanzella[1]
Christophe Verinaud[4]
Marco Xompero[1]
Simone Zaggia[1]

[1] INAF, Italy
[2] NUIG, Galway, Ireland
[3] CNRS/INSU, Grenoble, France
[4] ESO

**The Multi-conjugate Adaptive Optics RelaY (MAORY) is the adaptive optics (AO) module for the Extremely Large Telescope (ELT) that will provide two gravity-invariant ports with the same optical quality for two different client instruments. It will enable high-angular-resolution observations in the near-infrared over a large field of view (~ 1 arcminute$^2$) by real-time compensation of the wavefront distortions caused by atmospheric turbulence. Wavefront sensing is performed using laser and natural guide stars while the wavefront sensor compensation is performed by an adaptive deformable mirror (DM) in MAORY which works together with the telescope's adaptive and tip-tilt mirrors M4 and M5 respectively.**

## Introduction

MAORY will provide the ELT with two adaptive optics modes: the single-conjugate adaptive optics (SCAO) mode, which provides a very high correction over a field of view (FoV) of diameter ~ 10 arcseconds, with performance rapidly degrading with distance from the bright natural star used to probe the wavefront, and a multi-conjugate adaptive optics (MCAO) mode, which provides a moderate correction over a FoV of diameter ~ 60 arcseconds, with pretty homogeneous performance over the whole FoV.

MAORY is designed to support two different instruments, each with the same optical quality and with a gravity-invariant port. One of these two instruments will be the Multi-adaptive optics Imaging CamerA for Deep Observations (MICADO) near-infrared camera (Davies et al., 2018), while the second one is as yet undefined. The SCAO module is being developed within the MICADO consortium with contributions from MAORY and is described in Davies et al. (p. 17). The MAORY project is now in its Phase B stage and is progressing towards its Preliminary Design Review in early 2021.

## Science drivers

The scientific application of the SCAO mode will be limited by the need for a bright (approximately $V \leq 16$ magnitudes) star within few arcseconds of the scientific target, while the MCAO mode will make use of three natural guide stars (NGS) (with $H \leq 21.0$ magnitudes) to be found within an annular patrol field with an inner radius of ~ 40 arcseconds and an outer radius of ~ 160 arcseconds. The three NGS will allow us to correct low-order modes of the wavefront distortions, while the six laser guide stars (LGS) will be used to correct for high-order modes. This will make it possible to get AO-assisted observations over a large fraction of the sky, meeting the system specification for sky coverage (≥ 50% over the whole sky).

Coupled with MICADO, MAORY will enable the ELT to perform diffraction-limited observations in the near-infrared. In imaging mode MAORY + MICADO will provide an option with a wide FoV (50.5 × 50.5 arcseconds) at pixel scale of 4 milliarcseconds and a high-resolution option with a 1.5-milliarcsecond pixel scale over 19 × 19 arcseconds. This will represent a major step forward, with a significantly better spatial resolution than that of the Hubble Space Telescope (HST) and even the James Webb Space Telescope (which has a pixel scale ~ 30 milliarcseconds pixel$^{-1}$). Long-slit spectroscopy will be covered with two settings: a short slit of 0.84–1.48 μm, and a long slit of 1.48–2.46 μm (see Davies et al., p. 17 for a detailed description of the MICADO observing modes).

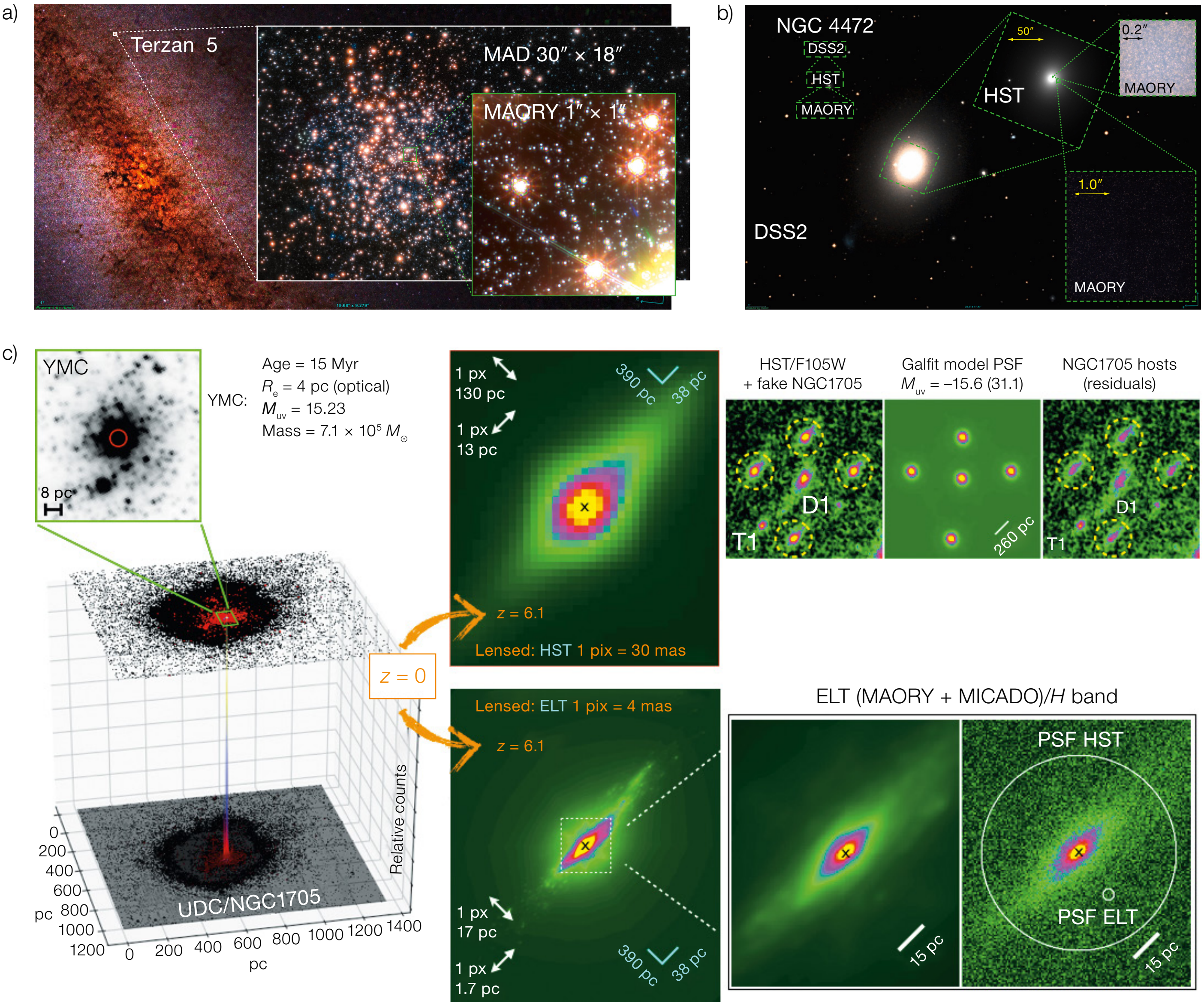

Figure 1. Combination of real and simulated images from the MAORY science cases White Book. a) Terzan 5 as imaged by MAD at VLT and by MAORY + MICADO. b) NGC 4470 as imaged by HST and MAORY + MICADO. c) 2D and 3D HST images of NGC 1705 and simulations at HST and MAORY + MICADO resolution lensed at $z = 6.1$.

The science cases for MAORY + MICADO have been widely explored by the MAORY science team. A preliminary collection of the cases studied is reported in the MAORY science cases White Book available on the MAORY website[1].

Together the science cases address many of the major questions in astrophysics:
– Planetary systems, including cases in our own Solar System, exoplanets and the formation of planetary systems;
– Nearby stellar systems, comprising stars and stellar systems within our own Galaxy and its satellites;
– The local Universe, with science cases aimed at studying the stellar content and the structure of distant stellar systems that can be at least partially resolved into individual stars. In many cases they will fall within the range of resolved systems only thanks to the advent of MAORY + MICADO;
– The high-redshift Universe, with the science cases addressing the formation of structures and cosmology using the formidable sensitivity and resolution of MAORY + MICADO to probe the very distant Universe and consequently the earliest phases of galaxy formation, as well as high-energy phenomena over the range of cosmic distance and time made accessible by the ELT.

In Figure 1 we show a collection of real and simulated images selected from the MAORY White Book to illustrate the extreme capability and versatility of the MAORY + MICADO combination. The top left panel shows a zoom-in on a candidate building block of the Galactic bulge, the globular cluster Terzan 5, as imaged by the Multi-conjugate Adaptive optics Demonstrator (MAD) at the VLT (real image) and by MAORY + MICADO. Thanks to its unprecedented angular resolution, MAORY + MICADO will make it possible to study the dynamics of stars very close to the centre of a globular cluster and to reveal the presence of an intermediate-mass black hole (IMBH) with a mass in the range $10^2$–$10^4$ $M_\odot$), shedding light on the origin of supermassive black holes (with masses larger than $10^6$ $M_\odot$). The top right panel shows a zoom-in on the giant elliptical galaxy NGC 4470 in Virgo, as imaged by a 2.5-m ground-based telescope, by the HST, and by MAORY + MICADO. Making use of spectroscopic and high-angular-resolution observations, we will be able to study the chemical composition of globular clusters from the Local Group to the VIRGO/Fornax galaxy cluster and the metallicity gradients in giant elliptical galaxies, and to resolve individual stars in nearby nuclear star clusters. The bottom panel illustrates a simulation of the ultra-compact dwarf galaxy NGC 1705 hosting its Super Star Cluster (SSC) (from Vanzella et al., 2019). On the left hand side is the HST image in 2D and 3D. To the right of that, the upper panel shows the modelled noiseless simulation at HST resolution lensed at $z$ = 6.1, and in the lower panel is the same simulation with MAORY + MICADO in the MCAO narrow-field mode, alongside a zoomed-in region in which the physical scale and the two point spread functions (HST and MAORY + MICADO) are indicated. This figure shows that, depending on the local magnification, a SSC at $z \sim 6$ will likely be resolved, and with a proper point spread function deconvolution we will resolve the light profile of the star cluster down to a resolution of 4–8 pc, allowing a proper photometric and spectroscopic analysis of SSCs. These kinds of studies will be fundamental tools with which to characterise star clusters at cosmological distances and their influence on the surrounding medium (for example, feedback and ionisation).

## Instrument design

### Optical design

The original optical design was altered in early 2020 in order to provide each supported instrument with a gravity-invariant port. The new MAORY main-path optical layout is envisaged to have eight reflections: two aspheric concave mirrors, two spherical DM, one convex and one concave, one dichroic and three fold mirrors. This optical layout (side and top view) is shown in Figure 2, where the right panel also shows the second port focal plane, the LGS module area envelope and the MICADO envelope. However, under the current technical specification, the MAORY baseline includes only one DM (with a convex shape and a diameter of about 900 mm) while the second, concave, DM is replaced by a rigid mirror that can itself be replaced with a DM in future. An aspheric correcting plate near the telescope focal surface (Rakich & Rogers, 2020a,b) allows the optical relay to simultaneously produce stigmatic images of the telescope focus and of laser guide stars over the full range of object distances. The plate correction also improves the image quality of system pupils and meta-pupils. In the presented optical configurations, the plates are at about 350 mm after the focal plane.

### Mechanical design

Starting from the new optical design, developed in 3D (not in a single plane as before; see Figure 2), the mechanical design has been completely revised. The main structure is now based on a latticework tower made of standard structural steel truss-beam-shaped pipes (welded and bolted) with different section properties. The instrument is completed by an LGS wavefront sensor (WFS) module with 6 beacons (upgradable to 8) arranged to form a 45-arcsecond asterism, an NGS WFS module with 3 low-order sensors and 3 references patrolling a technical field of 160 arcseconds, a real-time computer (RTC) sized for 8 LGS and 2 post-focal DM with 1500 actuators and a calibration and test unit.

A general overview of the design adopted as baseline is shown in Figure 3.

## Performance

The expected performance of the MAORY instruments is summarised in Figure 4. In the left panel we show the Strehl Ratio (SR) value as a function of the radial distance (in arcseconds) from the field centre for different atmospheric conditions, in the case of only one post-focal DM (as it is in the baseline configuration, solid lines) and in the case of two post-focal DM (dashed lines). In the right panel we show the sky coverage as a function of the SR value, again for different atmospheric conditions and with one DM (solid lines) and two DM (dashed lines). To calculate the sky coverage we make use of a statistical approach to calculate the number of times that a given performance is obtained on a simulated field at the south Galactic pole (SGP). The performance figures that we quote at 50% sky coverage are hence to be seen as the median performance obtained for a large set of random pointing at SGP which is indeed conservative with respect to the MAORY requirement to reach a SR of 30% on 50% of the observable sky (the goal being a SR of 50% with 2 DM).

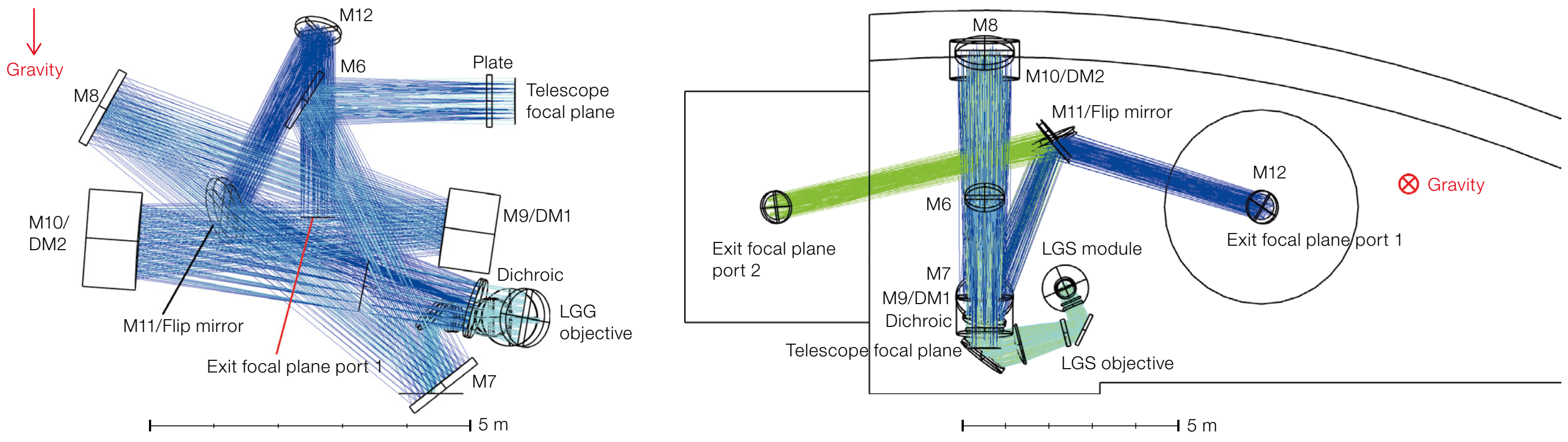

Figure 2. MAORY optical baseline side view layout (left) and top view layout (right).

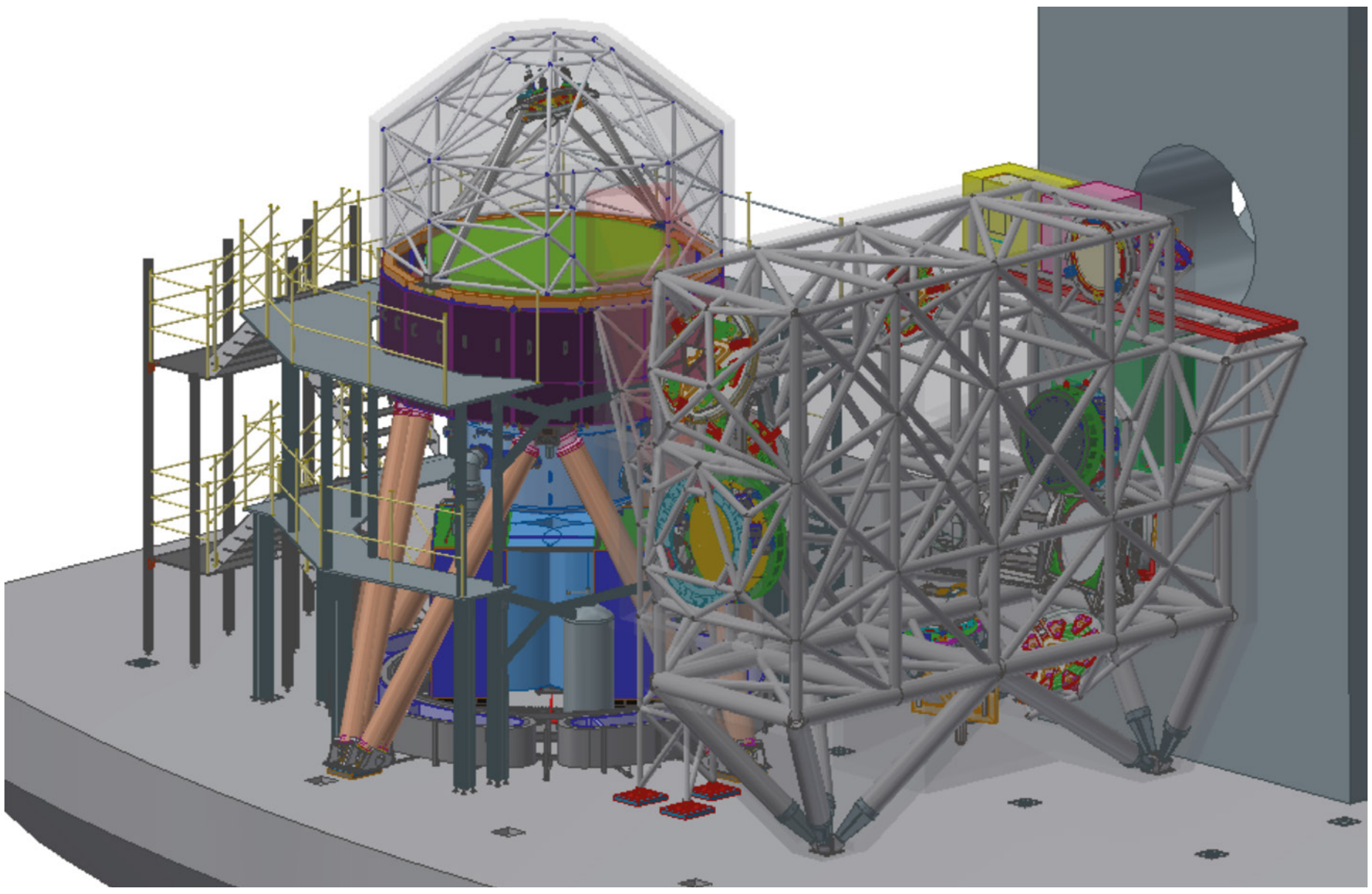

Figure 3. General overview of the MAORY instrument (with thermal cover shown transparent) installed on the Nasmyth platform with MICADO.

As shown in Figure 4, a system with a single DM is capable of delivering a SR above 40% at 50% of sky coverage — well above the requirement — while the presence of the second DM is fundamental to pushing the system towards maximal performance and higher robustness to varying atmospheric and observing condition.

Finally we would like to stress that a single set of performance figures cannot be fully representative of the real performance achievable on a specific target since many variable factors (like the atmospheric conditions, NGS asterism, sodium layer variability) play a role in the determination of the final result. As outlined also by Davies et al. (p. 17), potential users are encouraged to use the instrument data simulator ScopeSim[2] to familiarise themselves with the MAORY + MICADO instrument and to obtain a more accurate estimate of the performance on their own targets.

## Acknowledgements

MAORY is a consortium involving more than 70 people from institutes in Italy, France, Ireland and Germany working together with ESO[a]. The authors are grateful to their own institutes for financial support.

## Links

[1] MAORY web pages with the science cases White Book, a summary of the project and a list of people and partners involved: http://wwwmaory.oabo.inaf.it/
[2] ScopeSim is available at https://scopesim.readthedocs.io/

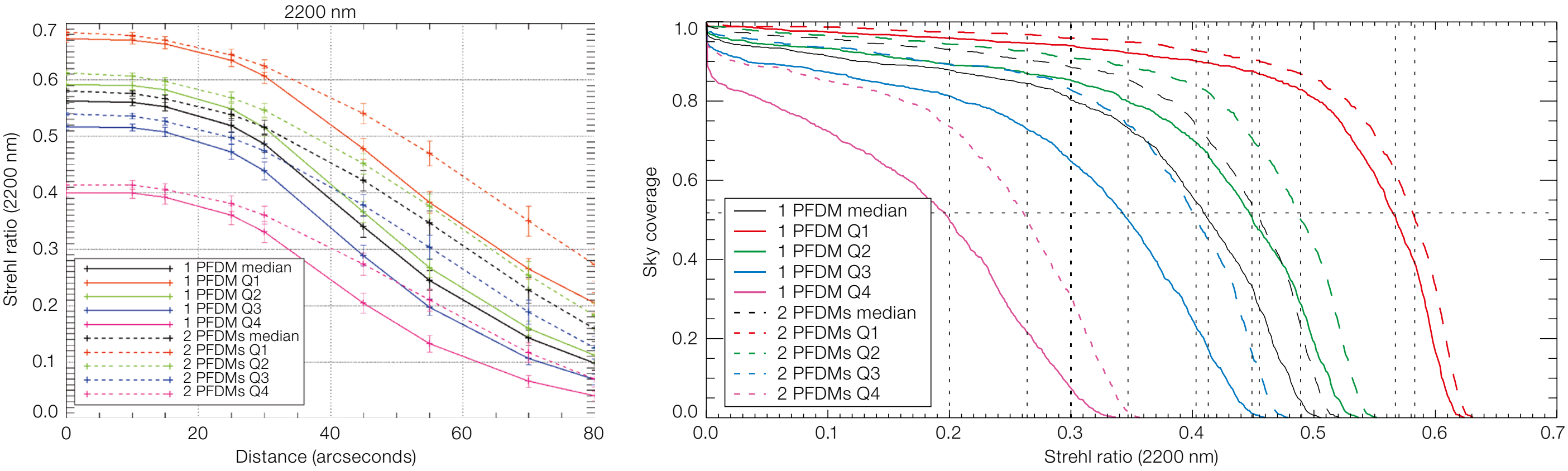

Figure 4. Summary of MAORY performance. Left: Strehl ratio (SR) as function of the radial distance from the field centre for different atmospheric conditions and 1 or 2 post-focal DM. Right: Sky coverage as function of the SR value for different atmospheric conditions and 1 or 2 post-focal DM.